\documentclass[twocolumn,showpacs,prl,amsmath,amssymb,showkeys]{revtex4-1}
\usepackage{epsfig}
\usepackage{rotating}
\usepackage{amssymb}
\usepackage{amsmath}
\usepackage{graphics,psfrag}
\usepackage{graphicx,psfrag}

\def\beq{\begin{eqnarray}}
\def\eeq{\end{eqnarray}}

\newcommand{\be}{\begin{equation}}
\newcommand{\ee}{\end{equation}}
\newcommand{\bea}{\begin{eqnarray}}
\newcommand{\eea}{\end{eqnarray}}

\def\eref#1{(\ref{#1})}

\def\eref#1{(\ref{#1})}
\def\bs{\boldsymbol}

\def\rd{\mathrm{d}}

\newcommand{\ti}{{t_{\mathrm{o}}}}
\newcommand{\tf}{{t_{\mathrm{f}}}}

\begin{document}

\title{Boundary layers in stochastic thermodynamics}

\author{Erik Aurell${}^{1,2,3}$}
\email{eaurell@kth.se}
\author{Carlos Mej\'ia-Monasterio${}^{4}$}
\email{carlos.mejia@upm.es}
\author{Paolo    Muratore-Ginanneschi${}^{5}$}
\email{paolo.muratore-ginanneschi@helsinki.fi}
\affiliation{${}^1$ACCESS Linnaeus Centre, KTH, Stockholm Sweden}
\affiliation{${}^2$Dept.~Computational Biology,  AlbaNova University Centre,
  106 91 Stockholm, Sweden}
\affiliation{${}^3$Aalto    University   School   of    Science,   Helsinki,
  Finland}
\affiliation{${}^4$Laboratory  of Physical  Properties, Department  of Rural
  Engineering, Technical  University of Madrid,  Av.  Complutense s/n,
  28040 Madrid, Spain}
 \affiliation{${}^5$University  of Helsinki,  Department of  Mathematics and
   Statistics P.O.  Box 68 FIN-00014, Helsinki, Finland}

\begin{abstract}
  We study the problem of  optimizing released heat or dissipated work
  in  stochastic  thermodynamics.    In  the  overdamped  limit  these
  functionals  have  singular  solutions,  previously  interpreted  as
  protocol jumps. We show that a regularization, penalizing a properly
  defined  acceleration, changes  the  jumps into  boundary layers  of
  finite width. We show that  in the limit of vanishing boundary layer
  width no heat is dissipated in the boundary layer, while work can be
  done.  We  further give  a new interpretation  of the fact  that the
  optimal  protocols in  the  overdamped limit  are  given by  optimal
  deterministic transport (Burgers equation).
\end{abstract}

\maketitle

With the advent of micromanipulation, thermodynamic quantities such as
work  and  heat  have  taken  new  operational  meaning  for  isolated
microstates or single trajectories in phase space. The problem is then
naturally posed to optimize such fluctuating quantities by varying the
externally imposed conditions, usually  called the protocol. The prime
experimental system  for such potentially  optimized micromanipulation
is  particles  or  molecules  in  optical  traps  \cite{BLR05,Astu07}.
Protocol optimization may  also turn out to be  important in improving
novel computational schemes harnessing the advances of non-equilibrium
statistical   physics    \cite{Nilmeier11}.    Many   functionals   of
fluctuating paths  may conceivably be  optimized, but two of  the most
natural  and important  are obviously  expected released  heat  to the
environment, and  expected dissipated work.  For  specific examples in
stochastic  thermodynamics, systems  described by  overdamped Langevin
equations, Schmiedl  and Seifert showed that  the optimizing protocols
have  discontinuities  \cite{SS07}.   Several  studies have  tried  to
assign a physical meaning to such infinitely fast transformations, and
even to  look for  an approximate process  that would be  amenable for
real experiments  \cite{G-MSS08,TE08,GD10}.  In a  recent contribution
using a Hamilton-Jacobi-Bellman approach  we showed that the solutions
of  these  examples  are  special  cases  of  a  more  general  scheme
connecting     optimal    protocols    to     optimal    deterministic
transport~\cite{Aurell2011}.   The  discontinuities  or jumps  in  the
protocols  are   generic,  and  can  be  understood   as  the  optimal
deterministic  transport proceeding  at constant  speed from  start to
finish~\cite{Aurell2011}.  The  infinitely fast transformations should
be  smoothened by  inertial  effects either  in  the system,  or in  a
physical model of the protocol.

In this Letter  we show that a regularization  by current acceleration
(a concept to be defined) allows for equally explicit solutions to the
problem  and  direct  investigations  of  the  corresponding  boundary
layers. We hence show from  the limit of regularized solutions that no
heat is released during the fast transformations.  In this work we use
extensively forward and backward derivatives of the stochastic process
as                developed               for               stochastic
quantization~\cite{Nelson1967,Nelson1985}.  As a  side  effect we  are
thus  also  able  to  derive  the  earlier  results  on  deterministic
transport in an alternative way.

The model we consider is the dynamics of the nonequilibrium transition
of  finite duration  $\Delta t  = \tf-\ti$  described by  the Langevin
equations in the overdamped limit
\begin{equation}
\label{wd:sde}
d\bs{\xi}_{t}=-\frac{\bs{b}_{t}}{\tau}dt
+\sqrt{\frac{2}{\tau\,\beta}}\,d\bs{\omega}_{t} \ ,
\end{equation}
with initial value $\bs{\xi}_{\ti}=\bs{x}_{\mathrm{o}}$, drift
$\bs{b}_{t} = \partial_{\bs{\xi}_{t}}V(\bs{\xi}_{t},t)$ and
${\bs{w}}_{t}$ a vector valued white noise with covariance $\langle
\dot{\bs{w}}_{t} \dot{\bs{w}}_{t'} \rangle = \delta(t-t')$, and
mobility $\tau^{-1}$.  $\bs{\xi}_{t}$ is an $\mathbb{R}^d$-valued
stochastic process indexed by the open time interval
$\mathbb{I}=[\ti,\tf]$.  During the transition the control potential
changes from $V(\bs{x},\ti)=U_{\mathrm{o}}(\bs{x})$ to
$V(\bs{x},\tf)=U_{\mathrm{f}}(\bs{x})$ and the probability density
$\rho(\bs{x},t)$ evolves according to the Fokker-Planck equation
\begin{equation} \label{eq:F-P}
 \partial_{t}\rho-\frac{1}{\tau}\partial_{\bs{x}}
 \cdot(\rho\,\partial_{\bs{x}}V)=
 \frac{1}{\beta\,\tau}\partial_{\bs{x}}^{2}\rho \ .
\end{equation}

Following  \cite{Sekimoto98},   an  energy  balance   for  the  single
stochastic  trajectories $\bs{\xi}_{t}$ of  these dynamics  yields the
so-called  stochastic thermodynamics.  Defining the  work done  on the
system during the time interval $\Delta t$ as
\begin{equation} \label{def:work}
\delta W = \int_{\ti}^{\tf}\partial_{t}V(\bs{\xi}_{t},t)~\rd t \ ,
\end{equation}
and the heat released by the system as
\begin{equation} \label{def:heat}
\delta Q =-\int_{\ti}^\tf \! \dot{\bs{\xi}_{t}}
\circ \partial_{\bs{\xi}_{t}}V(\bs{\xi}_{t},t)~\rd t \ ,
\end{equation}
then the balance $dU = \delta W - \delta Q$ resembles the first law of
thermodynamics  over the  time  interval $[\ti,\tf]$.   Note that  the
product in \eref{def:heat} must be defined in the Stratonovich sense.

We now introduce the notions  of the \textit{current velocity} and the
\textit{osmotic  velocity}   associated  to  the   stochastic  process
$\bs{\xi}_{t}$.   Assuming  that   \eref{wd:sde}  leads  to  a  smooth
diffusion  process  described  by  a  transition  probability  density
$p(\bs{x}_t,t|\bs{y}_s,s)$, the mean forward derivative is defined as
\begin{equation} \label{eq:Df}
D\bs{\xi}_t := \lim_{t^{\prime}\downarrow t}  \int\! d\bs{x}
\frac{\bs{x}-\bs{\xi}_t}{t^{\prime}-t}~ p(\bs{x},t^{\prime}|\bs{\xi}_t,t)
\equiv \frac{\bs{b}_t}{\tau} \ ,
\end{equation}
The  mean  backward derivative  can  be  written  similarly using  the
opposite      conditional      probability      $p(\bs{\xi}_t,t      |
\bs{x},t^{\prime})$.  For  Markov processes we can  use Bayes' formula
and write instead
\begin{eqnarray} \label{eq:Db}
D_*\bs{\xi}_t := \lim_{t^{\prime}\uparrow t}   \int\! d\bs{x}\frac{\bs{\xi}_t
 -\bs{x}}{t-t^{\prime}} \frac{p(\bs{\xi}_t,t|\bs{x},t^{\prime})
\rho(\bs{x},t^{\prime})}{\rho(\bs{\xi}_t,t)} \equiv \frac{\bs{b}_{*t}}{\tau} \ .
\end{eqnarray}
The mean forward and mean backward derivatives are related by
\begin{equation}
\bs{b}_{* t} = \bs{b}_t - \frac{2}{\beta\tau} \partial_{\bs{\xi_t}} 
\ln \rho(\bs{\xi}_t,t) \ ,
\end{equation}
and  the   current  velocity  $\bs{v}_t$  and   the  osmotic  velocity
$\bs{u}_t$ are
\begin{eqnarray}
\label{cv}
\bs{v}_t &=& (\bs{b}_{t} + \bs{b}_{* t})/2\tau \ , \\
\label{ov}
\bs{u}_t &=& (1/\beta\tau)  \partial_{\bs{\xi}_t}  \ln  \rho(\bs{\xi}_t,t) \ .
\end{eqnarray}
For any smooth function $f(\bs{\xi}_t)$ we have
\begin{equation}
\label{symder}
\left(\frac{D + D_*}{2}\right) f = \left(\partial_t + \bs{v}
\cdot\partial_{\bs{x}}\right) f \ ,
\end{equation}
while the mean  forward (or mean backward) derivative  by itself has a
diffusive  term, in  the symmetric  derivative of  Eq.~\ref{symder} it
cancels  out.  Correspondingly, the  Fokker-Planck equation  is always
deterministic mass transport in terms of the current velocity
\begin{equation} \label{eq:massevol}
\partial_{t}\rho+\partial_{x}\cdot(\rho\,\bs{v})=0 \ .
\end{equation}
We now use  the current velocity and osmotic velocity  in the heat and
work functionals  over the interval  $\mathbb{I}$, which we  define as
the  expectation  values  $\mathcal{W}  = \mathrm{E}~  \delta  W$  and
$\mathcal{Q}  = \mathrm{E}~  \delta Q$  respectively.  Straightforward
application  of the  It\^o lemma  (see e.g.\cite{Durrett})  yields the
heat functional
\begin{equation} 
\label{eq:oldQ}
\mathcal{Q}=\mathrm{E}\int_{t_{o}}^{t_{f}}
\left[d\bs{\xi}_{t}\cdot\bs{b}_{t}+
\frac{dt}{\beta\,\tau}\partial_{\bs{\xi}_{t}}
\cdot\bs{b}_{t}\right] \ .
\end{equation}
If the  probability measure $\rho$  decays sufficiently fast  after an
integration by parts we can write
\begin{equation}
\label{eq:Qvu}
\mathcal{Q}=\mathrm{E}\int_{t_{o}}^{t_{f}}dt
\left[\parallel\bs{v}_{t}\parallel^{2}
+\bs{u}_{t}\cdot\bs{v}_{t}\right] \ .
\end{equation}
Probability conservation and the definition of $\bs{u}$ then yield
\begin{equation} \label{result:1}
\mathcal{Q}=\frac{1}{\beta}\mathrm{E}\ln\frac{\rho_{\tf}}{\rho_{\ti}}
+\mathrm{E}\int_{\ti}^{\tf}dt\,\tau \parallel\bs{v}_{t}\parallel^{2} \ .
\end{equation}
From this follows immediately an inequality for the work:
\begin{eqnarray}
\label{eq:2ndlaw}
\mathcal{W}\geq \mathrm{E}\left\{U_{\mathrm{f}}-U_{\mathrm{o}}+\frac{1}{\beta}
\ln\frac{\rho_{\tf}}{\rho_{\ti}}\right\}=F_{\mathrm{f}}-F_{\mathrm{o}} 
\equiv \mathcal{F} \ ,
\end{eqnarray}
which is a form of the second law of thermodynamics.

In  our  earlier contribution~\cite{Aurell2011}  the  control was  the
drift $\bs{b}$, and the  functional was (\ref{eq:oldQ}). Proceeding as
above we can take the control to be $\bs{v}$, and the functional to be
(\ref{result:1}).  Given that  (\ref{symder})  and (\ref{eq:massevol})
are  already  inviscid  equations  this  means that  we  can  directly
interpret (\ref{result:1}) as a deterministic optimization problem the
solution of  which must  be an inviscid  equation (diffusion  does not
appear).    To  find   that  inviscid   equation,  which   is  Burgers
equation~\cite{Aurell2011}, $\bs{v}=\partial_x\psi/\tau$,
\begin{eqnarray}
\label{Burgers}
\partial_{t}\psi+\frac{\parallel\partial_{\bs{x}}\psi\parallel^{2}}
{2\tau}=0 \ ,
\end{eqnarray}
explicit calculations equivalent to those in~\cite{Aurell2011} must be
performed.  From (\ref{Burgers}) it follows that
\begin{eqnarray}
\mathrm{E}\int_{t_{o}}^{t_{f}}dt\,\tau \parallel\bs{v}_{t}\parallel^{2}=
2\,\mathrm{E}\int_{t_{o}}^{t_{f}}dt\,\frac{d \psi_{t}}{d t} \ ,
\end{eqnarray}
implying for  the heat released during the  optimal transformation the
expression \cite{Aurell2011}
\begin{eqnarray}
\mathcal{Q}_{\star}=\mathrm{E}\left\{2\,(\psi_{t_{f}}-\psi_{t_{o}})+
\frac{1}{\beta}\ln\frac{\rho_{\tf}}{\rho_{\ti}}\right\} \ .
\end{eqnarray}
with $\rho$  evolving according to \eref{eq:massevol}.  In the special
case     of     Gaussian      initial     and     final     densities,
$\rho\left(\bs{x},\ti\right)=(\beta/2\pi)^{d/2}
e^{-(\beta\parallel\bs{x}\parallel^{2}/2)}$                         and
$\rho\left(\bs{x},\tf\right)=(\beta/2\,\pi\,\sigma^{2})^{d/2}
e^{-(\beta\parallel\bs{x}-\bs{h}\parallel^{2}/2\,\sigma^{2})}$     with
$\bs{h}$ a constant vector,
the heat released by the optimal protocol over an a time horizon
$\Delta t$ is
\begin{eqnarray} \label{eq:Qoptim}
\mathcal{Q}_{\star}(\boldsymbol{h},\sigma)=
\frac{d}{2\,\beta}\ln\frac{1}{\sigma^{2}}
+\frac{\tau}{\Delta t}\left[\frac{d\,(\sigma\!-\!1)^{2} }{\beta}+
\parallel\bs{h}\parallel^{2}\right] \ ,
\end{eqnarray}
and the optimal current velocity 
\begin{eqnarray}
\label{heat-opt:vopt}
\bs{v}_{\star}\left(\bs{x},t\right)
=\frac{(\sigma-1)\,\bs{x}+\bs{h}}{\Delta t\,\sigma_{t}} \ ,
\end{eqnarray}
with $\sigma_t = [(\tf-t) + (t-\ti)\sigma]/\Delta t$ a linear function
of $t$.  A surprising property of this optimal driving (first obtained
in \cite{SS07}  for the minimization of the  work \eref{def:work}), is
the existence of discontinuities at the initial and final times of the
transformation.

We  will now  turn  to the  main topic  of  this Letter,  which is  to
regularize the optimization by penalizing the current acceleration
\begin{eqnarray}
\bs{a}_{t}=\left(\frac{D+D_{*}}{2}\right)\bs{v}_{t}
\end{eqnarray}
We  note  that  $\bs{v}$  and  $\bs{a}$  are  as  rough  functions  as
$\bs{\xi}$   along  trajectories   (but  no   rougher).   The  current
acceleration  would  be  a  complicated  expression in  terms  of  the
original  drift  field  $\bs{b}$  and  density  field,  but  the  heat
functional regularized by current acceleration preserves the same time
symmetry as the heat functional itself.  With these preliminaries, the
problem of  determining the minimal heat released  in a transformation
between given states  reached with assigned values of  the initial and
final current velocity  reduces to the problem of  finding the minimum
of the functional
\begin{eqnarray} \label{heat-opt:heat}
\lefteqn{
\mathcal{A}:=\mathrm{E}\int_{t_{o}}^{t_{f}}dt\,\tau\left(
\parallel\bs{v}_{t}\parallel^{2} +  \varepsilon\,\tau^{2}\, \parallel\bs{a}_{t}\parallel^{2}\right)}
\nonumber\\&&
\hspace{1.0cm}+ \mathrm{E}\int_{t_{o}}^{t_{f}}dt\,
\bs{\lambda}\cdot\left[\bs{v}_{t}-
\frac{\bs{\phi}(\bs{x}_{o})-\bs{x}_{o}}{\Delta t}\right]\hspace{2.0cm}
\end{eqnarray}
In \eref{heat-opt:heat} the Lagrange multiplier $\boldsymbol{\lambda}$
enforces the constraint
\begin{eqnarray}
\label{eq:bc-2}
\bs{x}_{\tf} = \bs{\phi}(\bs{x}_{o}) \ , \ \mathrm{with} \quad
\bs{x}_{\ti} = \bs{x}_{o} \ \mathrm{and} \quad \dot{\bs{x}}_{t}=\bs{v}_{t} \ .
\end{eqnarray}
and  the map  $\bs{\phi}$ specifies  the relation  between the  initial and
final states
\begin{eqnarray}
\rho_{\tf}(\phi(\bs{x}))\left|\det \frac{\partial 
\bs{\phi}(\bs{x})}{\partial \bs{x}}\right|= \rho_{\ti}(\bs{x}) \ .
\end{eqnarray}

\begin{figure}[!t]
\begin{center}
\psfrag{A}[b][b][1.2]{$\mathcal{A}_\star$}
{
  \includegraphics[width=0.95\linewidth]{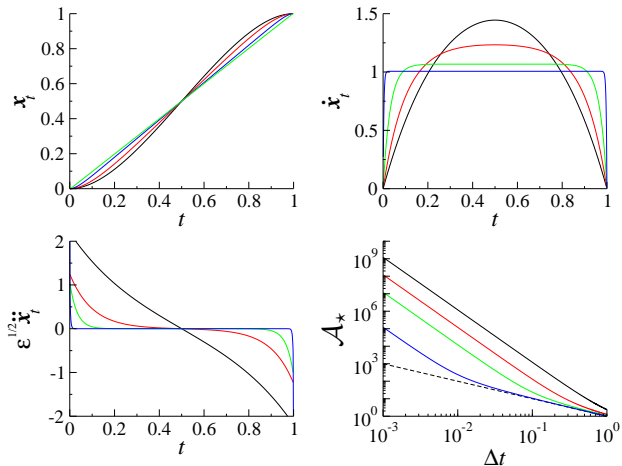}
}\caption{(Color  online)   Average  position  ${\bs{x}}_t$,  velocity
  $\dot{\bs{x}}_t$,          normalized         acceleration         $
  \sqrt{\varepsilon}\,\ddot{\bs{x}}_t$ as  obtained from \eref{sol-1},
  and  averaged  action   $\mathcal{A_*}$  \eref{action}  for  a  time
  interval   $[0,1]$  $\tau=h=\sigma=1$,   and  different   values  of
  $\varepsilon$:  $0.1$  (black curve),  $0.01$  (red curve),  $0.001$
  (green curve) and $0.00001$ (blue curve).  In the bottom-right panel
  the   dashed   line   corresponds   to  the   overdamped   value,
  $\varepsilon=0$. \label{fig-Q}}
\end{center}
\end{figure}
By (\ref{cv}), (\ref{ov}) we can write the initial current velocity as
\begin{equation}
\bs{v}_{\ti} = \frac{\bs{b}_{\ti}}{\tau} - \frac{1}{\beta\,\tau}
\partial_{\bs{x}}\ln \rho_{\ti}\left(\bs{x}\right) \ ,
\end{equation}
and  similarly for  $\bs{v}_{\tf}$.  It  follows that  if  the current
velocity vanishes at the boundary of the control horizon,
\begin{eqnarray}
\label{eq:bc-1}
\bs{v}_{\ti} = \bs{v}_{\tf}=0
\end{eqnarray}
the initial and final  probability densities correspond to equilibrium
states.  Furthermore, if the initial  an final states are Gaussian, as
used above  to obtain  \eref{eq:Qoptim}, then the  boundary conditions
\eref{eq:bc-2} reduce to
\begin{equation} \label{eq:bc-3}
\bs{x}_{\ti} = \bs{x}_{o} \ , \mathrm{and} \quad \bs{x}_{o} =
(\bs{x}_{\tf}-\bs{h})/\sigma \ .
\end{equation}
In general finding  the map $\bs{\phi}$ is the  main obstacle hindering the
derivation  of  explicit  solutions.   If we,  however,  consider  the
initial state and  the map $\bs{\phi}$ as boundary input  we can recast the
optimization problem into the simpler problem of minimizing the action
of   a  classical   unstable   oscillator  in   a  shifted   potential
$\tau    \parallel\bs{y}-\bs{\lambda}/(2\,\tau)\parallel^{2}$.     The
identifications  $\bs{y}_t=\dot{\bs{x}}_t =  \bs{v}_{t} $  provide the
connection  to  the  original  problem  and  the  boundary  conditions
\eref{eq:bc-1}  and \eref{eq:bc-3}.   From the  stationarity condition
for Eq.~\eref{heat-opt:heat}, we obtain the Euler-Lagrange equation
\begin{equation} \label{EL}
2\tau\left(\varepsilon\tau^2 \dddot{\bs{x}}_t -
    \dot{\bs{x}}_t\right) = \bs{\lambda} \ ,
\end{equation}
whence for  the boundary conditions  \eref{eq:bc-2}, \eref{eq:bc-1} it
follows
\begin{eqnarray}
\label{sol-1}
\dot{\bs{x}}_{t}
=\frac{\bs{\lambda}}{2\,\tau}\left[1-\frac{\cosh \left(\frac{2\,
(t-t_{o})-\Delta t} {2 \,\tau\,  \sqrt{\varepsilon }}\right)}
{\text{cosh}\left(\frac{\Delta t}{2 \,\tau\,  
\sqrt{\varepsilon }}\right)}\right] \ ,
\end{eqnarray}
with
\begin{equation}
\bs{\lambda}=-\frac{2\tau\,[\bs{h}+\left(\sigma-1\right)\,\bs{x}_{o}]}
{\Delta t-2 \,\tau\,  \sqrt{\varepsilon } 
\tanh \left(\frac{\Delta t}{2 \,\tau\, \sqrt{\varepsilon }}\right)} \ .
\end{equation}
The average  position ${\bs{x}}_t$ and  acceleration $\ddot{\bs{x}}_t$
are  obtained from  \eref{sol-1}. The  convergence of  the regularized
solution toward the overdamped  case of \cite{Aurell2011} is shown in
Fig.~\ref{fig-Q}.   Furthermore,   expressing  the  action  functional
\eref{heat-opt:heat} in terms of the stationary solution and averaging
over the initial state we obtain
\begin{eqnarray} \label{action}
\mathcal{A}_{\star}(\bs{h},\sigma,\varepsilon) = \frac{\tau\,
\parallel\bs{h}\parallel^{2}+d\,\left(\sigma-1\right)^{2}\beta^{-1}}
{\Delta t-2 \,\tau\,  \sqrt{\varepsilon } \tanh 
\left(\frac{\Delta t}{2 \,\tau\, \sqrt{\varepsilon }}\right)} \ ,
\end{eqnarray}
It  is  straightforward to  verify  that  in  the limit  of  vanishing
$\varepsilon$,   $E\ln   (\rho_{\tf}   /   \rho_{\ti})   /   \beta   +
\mathcal{A}_{\star}$    reduces   to    the    overdamped    result
\eref{eq:Qoptim} (see bottom-right panel of Fig.~\ref{fig-Q}).
\begin{figure}[!t]
\begin{center}
  \includegraphics[width=\linewidth]{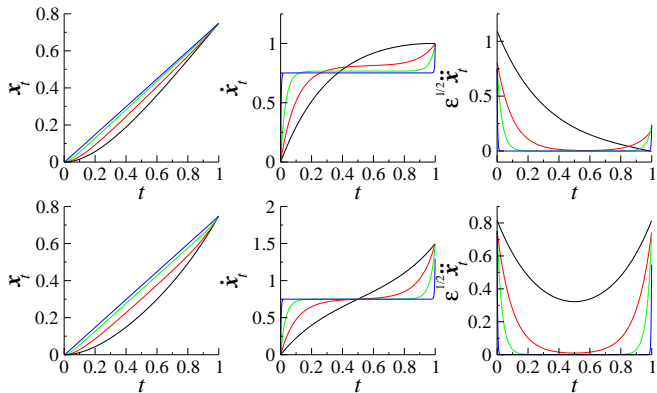}
  \caption{(Color  online)  Average  position  ${\bs{x}}_t$,  velocity
    $\dot{\bs{x}}_t$,              normalized             acceleration
    $\sqrt{\varepsilon}\,\ddot{\bs{x}}_t$     as     obtained     from
    \eref{vel-2}, for a time interval $[0,1]$, $x=-0.1$, $\sigma=3.5$,
    $\tau=h=1$, and  $y_{\tf}=1$ (upper panels),  $y_{\tf}=1.5$ (lower
    panels). The different curves  are for $\varepsilon$: $0.1$ (black
    curve), $0.01$  (red curve),  $0.001$ (green curve)  and $0.00001$
    (blue curve).  \label{fig-W}}
\end{center}
\end{figure}
Furthermore, for any small but finite $\varepsilon$ our regularization
unambiguously  determines   through  \eref{cv}, \eref{ov}  the  control
potential $V$ ($\boldsymbol{b}_{t}=\partial_{\boldsymbol{\xi}_{t}}V$)
in the closed  control interval $\mathbb{I}$. This means  that for any
$\varepsilon\,>\,0$ the optimal work expression
\begin{eqnarray}
\label{}
\mathcal{W}_{\star}=\mathrm{E}\left\{U_{\mathrm{f}}-U_{\mathrm{o}}+\ln\frac{\rho_{\tf}}
{\rho_{\ti}}\right\}+\mathcal{A}_{\star}
\end{eqnarray}
is well defined. In particular, for transformations between
equilibrium Gaussian sates we have immediately
$\mathcal{W}_{\star}=\mathcal{Q}_{\star}$.

Finally,  we consider the  minimization of  \eref{heat-opt:heat} under
the hypothesis that the final state is still Gaussian but \emph{out of
  equilibrium}.   In particular,  we suppose  the final  value  of the
control    potential    $U_{\mathrm{f}}    =   c\,|\boldsymbol{x}    -
\boldsymbol{h}  |^{2}/2$  to  differ  from the  osmotic  (equilibrium)
potential  $\ln  [\rho_{f}  \left(   2\,  \pi\,  \sigma^{2}  /  \beta
\right)^{d/2}]   =  -|  \boldsymbol{x}   -  \boldsymbol{\mu}   |^{2}  /
2\sigma^{2}$, thus implying  a non-vanishing final current velocity.
Proceeding as before we obtain
\begin{eqnarray} \label{vel-2}
\dot{\boldsymbol{x}}_{t}= 
G_{\bs{\lambda}}\sinh\left(\!\frac{t-\ti}{\sqrt{\varepsilon}\tau}\!\!\right)
+ \frac{\bs{\lambda}}{2\tau}\!\left[\!\cosh\left(\!\frac{t-\ti}
{\sqrt{\varepsilon}\tau}\!\!\right)\!-1\!\right] \ ,
\end{eqnarray} 
with       $G_{\bs{\lambda}}        =       [2\tau\bs{y}_{\tf}       -
\bs{\lambda}(c_{\bs{\lambda}}-1)] /2 \tau s_{\bs{\lambda}}$ and where,
to       ease       notation,      $s_{\bs{\lambda}}=\sinh\left(\Delta
  t/\sqrt{\varepsilon}\tau\right)$,             $c_{\bs{\lambda}}=\cosh
\left(\Delta t/\sqrt{\varepsilon}\tau\right)$, and
\begin{eqnarray}
\boldsymbol{\lambda}=\frac{c_{\bs{\lambda}}+1}
{s_{\bs{\lambda}}}
\frac{\bs{y}_{\tf}\sqrt{ \varepsilon}\tau(c_{\bs{\lambda}}\!-\!1)
-(\bs{\mu}+\bs{x}_{\mathrm{o}}\,(\sigma\!-\!1))s_{\bs{\lambda}}}
{(\Delta t/2\tau)(c_{\bs{\lambda}}+1) - \sqrt{\varepsilon} s_{\bs{\lambda}}} \ .
\end{eqnarray}
In the limit of vanishing regularization the minimal work done on the
system to operate the transformation tends to
\begin{eqnarray}
\label{minwork}
\mathcal{W}_{\star}\overset{\varepsilon\downarrow 0}{\to} \left(
\frac{c\,d\,\sigma^{2}}{2\,\beta} +\frac{c\,\parallel\boldsymbol{\mu}
-\boldsymbol{h}\parallel^{2}}{2}\right)
+\mathcal{Q}_{\star}(\bs{\mu},\sigma)
\end{eqnarray}
whilst \emph{within}  the open interval $(\ti,\tf)$ the  mean state of
the           system           changes           linearly           as
$\bs{x}_{t}=\boldsymbol{x}+(t-\ti)[\bs{\mu}+\bs{x}\,(\sigma-1)]/\Delta
t$  independently   of  the  final  value  of   the  current  velocity
$\bs{y}_{tf}$. We illustrate this phenomenon in Fig.~\ref{fig-W}.

From     \eref{minwork}    we     can    determine     the    Gaussian
\emph{nonequilibrium}  state  which,  given  the final  value  of  the
control  potential  $U_{f}$, minimizes  the  work.  A  straightforward
calculation    shows    that    the    minimum   is    attained    for
$\boldsymbol{\mu}=c\,\Delta  t\,\boldsymbol{h}/(c\,\Delta  t+2\,\tau)$
and      $\sigma^{-2}=[\sqrt{\Delta      t\,(c\,\Delta      t+2\,\tau)
  +\tau^{2}}-\tau]^{2}/(\Delta t)^{2}$. Thus, we recover the result of
\cite{SS07, Aurell2011}  for the minimal work  transforming an initial
equilibrium state under the constraint that the protocol at the end of
the  control  horizon  should  attain  an assigned  final  value.  Our
regularization  framework allows us  to interpret  such work  as lower
bound  over the work  done between  given states  positing that  it is
possible to retain  knowledge of the final protocol  but the knowledge
on the final non-equilibrium state is lost.

In  summary,  we  have  investigated  optimal  control  in  stochastic
thermodynamics.   First,  we  have  shown  that  the  optimal  control
equations for heat and  work transformations between given states have
a natural  interpretation in terms of functionals  definite under time
reversal   of  the   Markov   process  describing   the  overdamped
dynamics. Second, we have proposed a regularization framework in terms
of  current acceleration.   The regularization  allows us  to identify
without ambiguities the  internal energy of the system  with the drift
potential.   In the  limit  of vanishing  regularization, the  current
acceleration tends to zero within the control horizon but diverges (as
$\varepsilon^{-1/2}$    in   the    examples   considered)    at   the
control-horizon end-times  thus carrying  no contribution to  the heat
release.  Correspondingly,  the optimal protocol  converges toward the
overdamped solution by forming  boundary layers i.e. regions of faster
variation  at  the   control  horizon  boundaries.   As  $\varepsilon$
vanishes, these  regions shrink  to measure zero  sets over  which the
internal  energy forms  in the  limit discontinuities  bringing finite
contributions   to  the   work   done  on   the   system  during   the
transformation.    In  conclusion   we  achieved   a  fully-consistent
theoretical picture  of optimal overdamped  thermodynamics well suited
for the interpretation of experimental and numerical data.

This work  supported by the Swedish Research  Council through Linnaeus
Center ACCESS  and the FEDORA program  grant 129024 of  the Academy of
Finland (E.A.),  by the center  of excellence ``Analysis  and Dynamics
Research'' of the Academy  of Finland (P.M.G.). The authors gratefully
acknowledge the  hospitality of  NORDITA where part  of this  work has
been done during their stay  within the framework of the ``Foundations
and Applications of Non-Equilibrium Statistical Mechanics'' program.

\bibliography{reg-letter-5}%
\end{document}